\documentclass[aps,twocolumn,preprintnumbers,amsmath,amssymb,nofootinbib,superscriptaddress,notitlepage]{revtex4-1}

\usepackage{epsfig}
\usepackage[utf8]{inputenc}

\begin{document}

\title{Scale Invariance in Heavy Hadron Molecules}

	\author{Li-Sheng Geng}\email{lisheng.geng@buaa.edu.cn}
	\author{Jun-Xu Lu} 
	\author{M. Pavon Valderrama}\email{mpavon@buaa.edu.cn}
\affiliation{School of Physics and Nuclear Energy Engineering, \\
International Research Center for Nuclei and Particles in the Cosmos and \\
Beijing Key Laboratory of Advanced Nuclear Materials and Physics, \\
Beihang University, Beijing 100191, China} 

\date{\today}


\begin{abstract} 
\rule{0ex}{3ex}
We discuss a scenario in which the $P_c(4450)^{+}$ heavy pentaquark
is a $\Sigma_c \bar{D}^*$-$\Lambda_{c}(2595) \bar{D}$ molecule.
The $\Lambda_{c1} \bar{D} \to \Sigma_c \bar{D}^*$ transition
is mediated by the exchange of a pion almost on the mass shell
that generates a long-range $1/r^2$ potential.
This is analogous to the effective force that is responsible
for the Efimov spectrum in three-boson systems
interacting through short-range forces.
The equations describing this molecule exhibit approximate scale invariance,
which is anomalous and broken by the solutions.
If the $1/r^2$ potential is strong enough this symmetry survives in 
the form of discrete scale invariance, opening the prospect of an
Efimov-like geometrical spectrum in two-hadron systems.
For a molecular pentaquark with quantum numbers $\frac{3}{2}^{-}$
the attraction is not enough to exhibit discrete scale invariance,
but this prospect might very well be realized in a $\frac{1}{2}^{+}$ pentaquark
or in other hadron molecules involving transitions between particle
channels with opposite intrinsic parity and a pion near the mass shell.
A very good candidate is
the $\Lambda_{c}(2595) \bar{\Xi}_b - \Sigma_c \bar{\Xi}_b'$ molecule.
Independently of this, the $1/r^2$ force is expected to play
a very important role in the formation of this type of hadron molecule,
which points to the existence of
$\frac{1}{2}^{+}$ $\Sigma_c D^*$-$\Lambda_{c}(2595) D$ and
$1^+$ $\Lambda_{c}(2595) {\Xi}_b - \Sigma_c {\Xi}_b'$ molecules
and $0^{+}$/$1^{-}$ $\Lambda_{c}(2595) \bar{\Xi}_b - \Sigma_c \bar{\Xi}_b'$
baryonia.
\end{abstract}

\maketitle

The onset of scale invariance in two-body systems is a remarkable property.
It connects a series of seemingly disparate low-energy phenomena
in atomic, nuclear and particle physics under
the same theoretical description~\cite{Braaten:2004rn}.
When the scattering length $a_0$ of a two-body system is much larger
than any other scale, i.e. $a_0 \to \infty$, the system is invariant
under the scale transformation $r \to \lambda \, r$
with arbitrary $\lambda$~\cite{PavonValderrama:2007nu}. 
The low energy properties of this two-body system can be fully explained
independently of the underlying short-range dynamics.
That is,
few-body systems with a large scattering length admit a universal description.
Efimov discovered that three-boson systems exhibit a characteristic
three-body spectrum for $a_0 \to \infty$, where the binding energy
of the states is arranged in a geometric series~\cite{Efimov:1970zz}.
The continuous scale invariance of the three-body equations is anomalous
and the spectrum only shows discrete scale invariance
under the transformation $r \to \lambda_0 r$
where the value of $\lambda_0$ is now fixed.
Conversely if $E_n$ is the binding energy of a three-body state
there is another state with binding $E_{n+1} = E_n / \lambda_0^2$,
a prediction that was confirmed experimentally with Cs atoms
a decade ago~\cite{Kraemer:2006}.
This type of discrete geometrical spectrum also happens
in three-body systems containing at least
two-identical particles~\cite{Helfrich:2010yr},
or when the scattering is resonant
in higher partial waves~\cite{Helfrich:2011ut,Braaten:2011vf}.
This mechanism might be responsible for the binding of
the triton~\cite{Bedaque:1998kg}, $^4{\rm He}$~\cite{Konig:2016utl},
a series of halo
nuclei~\cite{Federov:1994cf,Horiuchi:2006ds,Canham:2008jd,Acharya:2013aea,Ji:2014wta}
and the Hoyle state~\cite{Hammer:2008ra,Higa:2008dn}.

There is a two-body system that is intimately related to the Efimov effect,
which is the $1/r^2$ potential~\footnote{While a two-body system with
infinite scattering length is scale invariant, a two-body
system with a $1/r^2$ potential can display at most
discrete scale invariance,
a possibility which will depend on the strength of the potential.
Continuous scale invariance for $1/r^2$ is broken by the existence of
a fundamental state with $E \neq 0$. Equivalently,
as happens in the three-boson system~\cite{Bedaque:1998km},
the renormalization of $1/r^2$ is non-trivial and requires
the appearance of a new energy scale and henceforth
that scale invariance is broken.}.
At zero energy the reduced Schr\"odinger equation for the s-wave becomes
\begin{eqnarray}
\label{eq:schro}
-u''(r) + \frac{g}{r^2}\,u(r) = 0 \, ,
\end{eqnarray}
which is obviously scale invariant
(for a finite energy analysis we refer to~\cite{Bawin:2003dm}).
The connection with the three-body system is apparent when one realizes
that it also contains a similar equation with an effective $1/\rho^2$
potential in the hyperradius $\rho$~\cite{PhysRevLett.71.4103}.
For $g > -1/4$ the equation above admits power-law solutions of the type
\begin{eqnarray}
u(r) = c_{+} r^{\frac{1}{2} + \nu} + c_{-} r^{\frac{1}{2} -\nu} \, ,
\end{eqnarray}
with $c_{\pm}$ constants and $\nu = \sqrt{1/4+g}$,
where scale invariance is lost.
For $g < -1/4$ we have instead solutions of the type
\begin{eqnarray}
u(r) = c\,r^{1/2} \sin{(\nu \log{\Lambda_2 r})} \, ,
\label{eq:zero}
\end{eqnarray}
with $c$ a constant, $\nu = \sqrt{-1/4 - g}$ and
$\Lambda_2$ an energy scale that depends on the short-range physics 
(it can be obtained from the energy of the fundamental state).
$\Lambda_2$ is the reason why exact scale invariance is broken and
its appearance resembles dimensional
transmutation~\cite{Camblong:2000qn,Camblong:2000ax}.
Now the solutions display discrete scale invariance with $r \to \lambda_0 r$,
where $\lambda_0 = e^{\pi / \nu}$~\cite{Braaten:2004pg,Hammer:2005sa}.
In turn there is a geometric bound state spectrum
where $E_{n+1} = E_{n} / \lambda_{0}^2$,
with $E_n$ and $E_{n+1}$ the energy of two consecutive states.
Incidentally this is a rare example of an anomaly
in quantum mechanics~\cite{Coon:2002sua}.
Here we make the observation that the $1/r^2$ potential can appear in
heavy hadron molecules, for instance the $P_c(4450)^{+}$
heavy pentaquark if it happens to be molecular
(the only other known example of a $1/r^2$ potential is the atom-dipole
interaction~\cite{Camblong:2001zt}).
There might be other two-hadron systems where the potential
might be attractive enough to exhibit discrete scale invariance.
The ideas presented here involve long-range physics and hence only apply
to molecular hadrons (i.e. non-relativistic bound states of two-hadrons)
that fulfill a series of conditions, but not to compact hadrons.

\begin{figure}[ttt]
\begin{center}
\includegraphics[width=9.5cm]{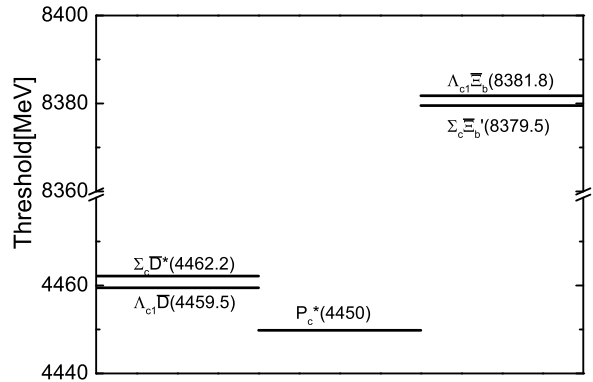}
\end{center}
\caption{
Location of the thresholds for the two scale invariant molecule candidates
considered in this work, the $\Lambda_{c1} \bar{D}$-$\Sigma_c \bar{D}^*$
and the $\Lambda_{c1} \bar{\Xi}_b$-$\Sigma_{c} \bar{\Xi}_b'$. We also
show the location of the $P_c^*$ for comparison.
}
\label{fig:threshold}
\end{figure}

The heavy pentaquarks $P_c(4380)^{+}$ and $P_c(4450)^{+}$,
$P_c$ and $P_c^*$ from now on, were discovered by the LHCb~\cite{Aaij:2015tga}
and are a recent and interesting addition to a growing family of
exotic hidden charm (and bottom) hadrons
that began with the $X(3872)$ more than ten years ago~\cite{Choi:2003ue}.
There is still a lot of discussion regarding the nature of
the $P_c$ and $P_c^*$, from the role of threshold
effects~\cite{Guo:2015umn,Mikhasenko:2015vca,Liu:2015fea,Meissner:2015mza},
to baryocharmonia~\cite{Kubarovsky:2015aaa},
a compact pentaquark~\cite{Diakonov:1997mm,Jaffe:2003sg,Yuan:2012wz,Maiani:2015vwa,Lebed:2015tna,Li:2015gta},
a heavy baryon-antimeson molecule~\cite{Chen:2015loa,Chen:2015moa,Roca:2015dva,He:2015cea,Xiao:2015fia} 
or other more exotic possibilities~\cite{Mironov:2015ica,Scoccola:2015nia}.
The $P_c^*$ is an interesting molecular candidate because its width is
not particularly big $\Gamma = 35 \pm 5 \pm 19\,{\rm MeV}$ and 
its closeness to the $\Sigma_c \bar{D}^*$ threshold,
see Fig.~(\ref{fig:threshold}).
As a matter of fact a series of works predicted the possibility of
a heavy baryon-antimeson molecule before the discovery
of the $P_c^*$~\cite{Wu:2010jy,Yang:2011wz,Xiao:2013yca,Karliner:2015ina}.
The probable quantum numbers of the $P_c$ and $P_c^*$ are $\frac{3}{2}^{-}$
and $\frac{5}{2}^{+}$ respectively, followed by $\frac{5}{2}^{+}$ and
$\frac{3}{2}^{-}$.
The standard molecular explanation for the $P_c^*$ heavy pentaquark is
that of a $\Sigma_c \bar{D}^*$ bound state, which prefers the quantum
number ${\frac{3}{2}}^{-}$ for the $P_c^*$.
Here we discuss the scenario in which the molecular $P_c^*$ also contains
a $\Lambda_{c1} \bar{D}$ component in addition to $\Sigma_c \bar{D}^*$,
where $\Lambda_{c1}$ denotes the $\Lambda_{c}(2595)$.
Burns~\cite{Burns:2015dwa} proposed this idea on the analogy between
the $D \bar{D}^* + D^* \bar{D}$ and the $Y_c^* \bar{D} + Y_c \bar{D}$ systems,
i.e. the X(3872) and the $P_c^*$, where $Y_c$, $Y_c^*$ are charmed baryons.
He argued that the most natural analog to the $D \bar{D}^* + D^* \bar{D}$
system is $\Sigma_c \bar{D}^* - \Lambda_{c1} \bar{D}$ on the basis that
the mass difference of the $\Lambda_{c1}$ and $\Sigma_c$ is very close
to the $D^*$ and $D$ splitting.
Here we will explore this possibility.

The low-energy dynamics of $\Sigma_c \bar{D}^* - \Lambda_{c1} \bar{D}$
is driven by one pion exchange (OPE) and is fascinating for two reasons.
First, if the $\Lambda_{c1} \bar{D}$ pair exchanges a pion to become
a $\Sigma_c \bar{D}^*$ pair the pion will be almost on the mass shell,
leading to a unusual long-range potential for strong interactions.
Second, the intrinsic parities of $\Lambda_{c1}(\frac{1}{2}^{-})$ and
$\Sigma_{c}(\frac{1}{2}^{+})$ are different while the ones
for the $D$ and $D^*$ are the same.
As a consequence OPE will switch odd (even) waves in the $\Lambda_{c1} \bar{D}$
channel to even (odd) waves in the $\Sigma_c \bar{D}^*$ one.
That is, there is a vector force: the analogous to the tensor one, except
that it carries orbital angular momentum $L=1$ instead of $L=2$.
The tensor force behaves as $1/r^3$ for $m_{\pi} r < 1$,
while the vector force as $1/r^2$.
This short-range property becomes long-range if the pion is near the mass shell.

We can compute the $\Sigma_c \bar{D}^* \to \Lambda_{c1} \bar{D}$ potential
from the heavy baryon chiral lagrangian of Cho~\cite{Cho:1994vg}
\begin{eqnarray}
\langle \Lambda_{c1} \bar{D} 
|  V_{\rm OPE}(\vec{r}) & | 
\Sigma_c \bar{D}^* \rangle =
\omega_{\pi}\,\tau\,\vec{\epsilon} \cdot \hat{r}\,\,W_E(r) \, ,
\label{eq:OPE}
\end{eqnarray}
with $\omega_{\pi} = m(\Lambda_{c1}) - m(\Sigma_c)$ the energy of the pion,
$\tau$ an isospin factor such that $\tau = \sqrt{3}$
for $I=1/2$ and $\tau = 0$ for $I=3/2$ and $\vec{\epsilon}$
the polarization vector of the incoming $\bar{D}^*$ meson.
$W_E$ reads
\begin{eqnarray}
W_E(r) = \frac{g_1 h_2\,\mu_{\pi}^2}{4\pi\,\sqrt{2}\,f_{\pi}^2}\,
\frac{e^{-\mu_{\pi} r}}{\mu_{\pi} r}\,\left( 1 + \frac{1}{\mu_{\pi} r} \right)
\, ,
\label{eq:vector}
\end{eqnarray}
with $g_1$ the axial coupling for the heavy mesons, $h_2$ the coupling
for the $\pi \Lambda_{c1} \Sigma_c$ vertex,
$f_{\pi} \simeq 130\,{\rm MeV}$ the pion decay constant and
$\mu_{\pi} = m_{\pi}^2 - \omega_{\pi}^2$
is the effective pion mass.
Besides, there is standard-range OPE in the $\Sigma_c \bar{D}^*$ channel 
while OPE vanishes in the $\Lambda_{c1} \bar{D}$ channel.

Actually $|\mu_{\pi}| \simeq 5-35\,{\rm MeV} \ll m_{\pi}$
depending on whether we exchange a charged or neutral pion, i.e.
the $\Sigma_c \bar{D}^* \to \Lambda_{c1} \bar{D}$
transition potential dominates the long-range dynamics
of the system for $1 / m_{\pi} < r < 1 / |\mu_{\pi}|$
(which is also the region of validity of the equations we will write below).
We stress that scale invariance is only approximate and broken by
two interrelated factors: (i) the pion is not exactly on the mass shell,
(ii) the  $\Sigma_c \bar{D}^*$ and $\Lambda_{c1} \bar{D}$
thresholds are a pair of ${\rm MeV}$ away from each other.
For the moment we will assume $\mu_{\pi} = 0$, which implies overlapping
thresholds.
In principle the widths of the $\Sigma_c$ and $\Lambda_{c1}$ baryons
are another factor to consider.
Yet the widths can be ignored if the time required for the formation of
the state is shorter than the life-time of its components: $\Gamma \ll m$,
with $\Gamma$ the width of the component and $m$ the mass of
the exchanged particle~\cite{Guo:2011dd}.
For the $\Sigma_c$ and $\Lambda_{c1}$ the widths are about a pair of MeVs,
well below $m_{\pi} \sim 140\,{\rm MeV}$.
The potential in the $I=1/2$ channel reads
\begin{eqnarray}
\langle \Lambda_{c1} \bar{D} 
| V_{\rm OPE}(\vec{r}) & |  \Sigma_c \bar{D}^* \rangle =
\frac{g_1 h_2\,\omega_{\pi}}{4\pi\,\,f_{\pi}^2}\,\sqrt{\frac{3}{2}}\,
\frac{\vec{\epsilon} \cdot \hat{r}}{r^2} + \mathcal{O}(\mu_{\pi}^2 r^2)\, ,
\nonumber \\
\end{eqnarray}
i.e. the $\mu_{\pi} = 0$ limit of Eqs.~(\ref{eq:OPE}) and (\ref{eq:vector}).
If we consider $J^P = \frac{3}{2}^{-}$
(the standard quantum numbers for a molecular pentaquark),
the partial waves contributing are $\Sigma_c \bar{D}^* ({}^2D_{3/2})$,
$\Sigma_c \bar{D}^* ({}^4S_{3/2})$,
$\Sigma_c \bar{D}^* ({}^4D_{3/2})$ and
$\Lambda_{c1} \bar{D} ({}^2P_{3/2})$.
In this partial wave basis the reduced Schr\"odinger equation
at zero energy reads
\begin{eqnarray}
- {\bf u}'' +
\left[ 2 \mu_{P_c^*} {\bf V}_{\rm OPE} + \frac{{\bf L}^2}{r^2} \right]
\,{\bf u} = 0 \, ,
\end{eqnarray}
where ${\bf u}$ is the wave function in vector notation and $\mu_{P_c^*}$
the reduced mass of the molecule (actually there is one reduced mass
for each particle channel, but here we can take their geometric mean).
The combination of the vector OPE potential and the centrifugal barrier reads
\begin{eqnarray}
2\mu_{P_c^*} {\bf V}_{\rm OPE} + \frac{{\bf L}^2}{r^2} &=&
\frac{{\bf g}({\frac{3}{2}}^{-})}{r^2}  \nonumber \\
&=& \frac{1}{r^2}
\begin{pmatrix}
6 & 0 & 0 & g \\
0 & 0 & 0 & g \\
0 & 0 & 6 & -g \\
g  & g  & -g  & 2
\end{pmatrix} \, .
\end{eqnarray}
That is, we have a four channel version of Eq.~(\ref{eq:schro}).
We can diagonalize the matrix ${\bf g}({\frac{3}{2}}^{-})$, in which case
we end up with four equations of the type
\begin{eqnarray}
- u_{i}'' + \frac{g_i}{r^2}\,u_i = 0 \, ,
\end{eqnarray}
where the $g_i$'s ($i = 1, 2, 3, 4$) are
the eigenvalues of ${\bf g}({\frac{3}{2}}^{-})$.
There are three positive and one negative eigenvalue
\begin{eqnarray}
g_i = \{ 6, 2, 3 + \sqrt{9 + 3 g^2}, 3 - \sqrt{9 + 3 g^2} \} \, , 
\end{eqnarray}
where the negative one can trigger discrete scale invariance.
This will happen if $| g | > 5 / (4 \sqrt{3}) \simeq 0.7217$.
However the value of $g$ for the $P_c^*$ molecule is
$g = 0.60^{+0.10}_{-0.10}\,h_2$, where we have used
$g_1 = 0.59 \pm 0.01 \pm 0.07$ from $D^* \to D \pi$
and $D^* \to D \gamma$ decays~\cite{Ahmed:2001xc,Anastassov:2001cw}.
This requires $| h_2 | > 1.21^{+0.25}_{-0.19} $, which is well above
$h_2 = 0.60 \pm 0.07$ from CDF~\cite{Aaltonen:2011sf} or
$h_2 = 0.63 \pm 0.07$ from the analysis of Ref.~\cite{Cheng:2015naa},
where in both cases $h_2$ is extracted from
$\Gamma(\Lambda_{c1} \to \Sigma_c \pi)$.
That is, there is not enough attraction to achieve discrete scale invariance.
For the $\frac{1}{2}^{-}$ molecule the matrix is different but the
attractive eigenvalue is still $g_{-}(\frac{1}{2}^{-}) = 3 - \sqrt{9 + 3 g^2}$,
requiring $| g | > 5 / (4 \sqrt{3})$.
The most interesting pentaquark-like molecule is the $\frac{1}{2}^{+}$,
with partial waves $\Sigma_c \bar{D}^* ({}^2P_{1/2})$,
$\Sigma_c \bar{D}^* ({}^4P_{1/2})$ and $\Lambda_{c1} \bar{D} ({}^2S_{1/2})$,
where
\begin{eqnarray}
{\bf g}({\frac{1}{2}}^{+}) &=& 
\begin{pmatrix}
2 & 0 & g \\
0 & 2 & -\sqrt{2} g \\
g  & -\sqrt{2} g & 0
\end{pmatrix} \, .
\end{eqnarray}
The attractive eigenvalue is $g_{-}(\frac{1}{2}^{+}) = 1 - \sqrt{1+ 3 g^2}$,
which requires $| g | > \sqrt{3} / 4 \simeq 0.4330$ and
$|h_2| > 0.73^{+0.11}_{-0.06}$, i.e. overlapping with
current estimations of $h_2$.
Finally the $\frac{3}{2}^{+}$, $\frac{5}{2}^{+}$ and $\frac{5}{2}^{-}$ cases
require $| g | > 7 \sqrt{3} / 4$, $| g | > 7 \sqrt{3} / 4$
and $| g | > 15 \sqrt{3} / 4$.
That is, the strength of the vector force is in general too weak
in the pentaquark-like molecules to achieve discrete scale
invariance, with the notable exception of $\frac{1}{2}^{+}$
which lies on the limit.

Yet the $P_c^*$ is not the only system where this can happen.
The general conditions for a $H_1 H_2 - H_1' H_2'$ hadronic molecule
to have scale invariance are: (i) the hadrons are particularly long-lived,
(ii) the mass difference of the hadrons
in each vertex is similar to that of a pseudo-Goldstone boson
$m(H_1)' - m(H_1) \simeq m(H_2) - m(H_2)' \simeq m_P$
(iii) the intrinsic parity of $H_2$ and $H_2'$ is the same,
while that of $H_1$ and $H_1'$ is different and
(iv) $H_1$ and $H_1'$ have the same spin for the pseudo-Goldstone
boson to be emitted in s-wave.
%
This applies as well if we substitute hadrons by antihadrons in one of
the vertices: the vector force will change sign but the eigenvalues
of the $1/r^2$ potential matrix will remain the same.
Notice that it is not strictly necessary to exchange a pion near the mass-shell
to have a long-range $1/r^2$ force. A kaon near the mass-shell
will also generate this type of force.

If we have the $\Lambda_{c1}$-$\Sigma_{c}$ on the one side,
besides the $D$-$D^*$, the $\Xi_b$-$\Xi_b'$ bottom baryon combination
also fulfills the previous conditions,
see Fig.~(\ref{fig:threshold}) for the threshold location.
In this regard
the $\Lambda_{c1} \bar{\Xi}_{b}$-$\Sigma_{c} \bar{\Xi}_{b}'$ system 
seems to be the best candidate for a scale invariant molecule
in the heavy sector.
The $\Lambda_{c1} \bar{\Xi}_{b} \to \Sigma_{c} \bar{\Xi}_{b}'$
potential for $I=1/2$ reads
\begin{eqnarray}
\langle \Sigma_c \bar{\Xi}_{b}' | 
V_{\rm OPE}(\vec{r}) & | \Lambda_{c1} \bar{\Xi}_b \rangle =
\frac{g_3 h_2\,\omega_{\pi}}{8\pi\,\,f_{\pi}^2}\,
\frac{\sigma_2 \cdot \hat{r}}{r^2} + \mathcal{O}(\mu_{\pi}^2 r^2) \, ,
\end{eqnarray}
where $g_3$ is the axial coupling for the $\bar{\Xi}_{b}' \bar{\Xi}_{b} \pi$
vertex and $\sigma_2$ is the Pauli matrix for that vertex.
If we consider states in which the $\Sigma_{c} \bar{\Xi}_{b}'$ is in s-wave
or alternatively in p-wave where the tensor force is attractive, we have
\begin{eqnarray}
0^{+} &=& \Sigma_{c} \bar{\Xi}_{b}' ({}^3P_0) - 
\Lambda_{c1} \bar{\Xi}_{b} ({}^1S_0)  \, , \\
0^{-} &=& \Sigma_{c} \bar{\Xi}_{b}' ({}^1S_0) - 
\Lambda_{c1} \bar{\Xi}_{b} ({}^3P_0) \, , \\
1^{-} &=& \Sigma_{c} \bar{\Xi}_{b}' ({}^3S_1-{}^3D_1) -
\Lambda_{c1} \bar{\Xi}_{b} ({}^1P_1-{}^3P_1) \, .
\end{eqnarray}
In these partial wave bases the ${\bf g}$ matrices read
\begin{eqnarray}
{\bf g}(0^{+}) &=&
\begin{pmatrix}
2 & g \\
g & 0 
\end{pmatrix} \, , \\
{\bf g}(0^{-}) &=&
\begin{pmatrix}
0 & g \\
g & 2 
\end{pmatrix} \, , 
\end{eqnarray}
\begin{eqnarray}
{\bf g}(1^{-}) &=&
\begin{pmatrix}
0 & 0 & \frac{1}{\sqrt{3}}\,g & -\sqrt{\frac{2}{3}}\,g \\
0 & 6 & -\sqrt{\frac{2}{3}}\,g & -\frac{1}{\sqrt{3}}\,g \\
\frac{1}{\sqrt{3}}\,g & -\sqrt{\frac{2}{3}}\,g & 2 & 0 \\
-\sqrt{\frac{2}{3}}\,g & -\frac{1}{\sqrt{3}}\,g & 0 & 2 
\end{pmatrix} \, .
\end{eqnarray}
For $| g  | > 3/4$ the attractive eigenvalue of the matrices above
will trigger discrete scale invariance.
The evaluation of $g$ depends on the axial coupling $g_3$,
which can be extracted from the $\Sigma_c \to \Lambda_c \pi$ decay,
yielding $g_3 = 0.973^{+0.019}_{-0.042}$~\cite{Cheng:2015naa}.
This translates into $g = 1.12^{+0.03}_{-0.05}\,h_2$
requiring $|h_2| > 0.67^{+0.03}_{-0.02}$, which is within the error of 
$h_2 = 0.63 \pm 0.07$~\cite{Cheng:2015naa}.

The approximate scale invariance of the Schr\"odinger equation
describing these hadronic molecules has long- and short-range consequences,
where the former --- the appearance of a geometric spectrum ---
depends on how far these systems are from $\mu_{\pi} = 0$.
For $\mu_{\pi} \neq 0$ scale invariance holds for 
\begin{eqnarray}
R_s < r < \frac{1}{|\mu_{\pi}|} \, ,
\end{eqnarray}
with $R_s$ the short-range scale, $1/m_{\pi}$ in this case~\footnote{
On a related note, a purely imaginary $\mu_{\pi}$ triggers
a repulsive correction
at second order perturbation theory. The effect is small
and is suppressed as $| \mu_{\pi} r |^3 / 3$ in the scale invariant region.}
(this also applies to three-boson systems after
the substitution $\mu_{\pi} \to 1/a_0$).
The existence of a geometric excited state requires the relative size of
the scale invariant window to be bigger than the discrete scaling factor.
For $P_c^*$-like molecules this window is $1/(R_s \mu_{\pi}) \sim 10 - 20$,
requiring the coupling $|g_{-}|$ to be about $1$ at least,
which is considerably larger than $1/4$.
That is, the observation of geometric states in hadron and atomic physics
share a similar difficulty: the fine-tuning of the pion mass (hadrons)
or the scattering length (atoms).
For atoms near a Feschback resonance this is solved
with a magnetic field~\cite{Braaten:2004rn}.
The equivalent for hadrons will be to fine-tune the pion mass in the lattice.
There is also the possibility of increasing $|g_{-}|$,
for instance by having a larger reduced mass (i.e. two bottom hadrons)
or if the exchanged particle is a kaon.
This can happen naturally in the heavy sector where there are still
plenty of hadrons to be discovered, of which a few might
be candidates for a long-range vector force.

Concerning the short-range consequences,
even if the vector force is not enough to trigger discrete scale invariance
it will still play a remarkable role in binding.
This is indeed analogous to the conjectured importance of Efimov physics
in light nuclei~\cite{Konig:2016utl} (despite the glaring absence of
Efimov states).
If the binding mechanism is s-wave short-range attraction,
a way to see this is the following:
for $r \leq R_s$ we will assume that OPE is not valid
and model the short-range interaction
with a delta-shell:
\begin{eqnarray}
V(r) = V_{\rm OPE}(r)\,\theta(r-R_s) +
\frac{C_0(R_s)}{4\pi R_s^2} \delta(r-R_s) \, ,
\end{eqnarray}
where $R_s$ is the short-range radius.
Then we calculate the relative strength of the coupling $C_0$ required
to have a bound state at zero energy in the presence/absence of a vector force.
In the one-channel problem of Eq.~(\ref{eq:schro}) for $g > -1/4$
and in the absence of tensor OPE,
the relative strength of $C_0$ is $(1/2 + \nu)$
of that required to bind if $g = 0$ (for $\mu_{\pi} R_s < 1$),
while for $g < -1/4$ it always binds (for $\mu_{\pi} = 0$).
Owing to scale invariance this happens independently of $R_s$.
Thus if $\nu \to 0$ ($g \to -1/4$) the short-range potential only has to be
half the normal strength to be able to bind the system.
If there is standard-range OPE or other intermediate range physics
this binding enhancement will change.
Taking $R_s = 1\,{\rm fm}$, $\mu_{\pi} = 0$ and $h_2 = 0.63$,
the $\Sigma_c \bar{D}^*$-$\Lambda_{c1} \bar{D}$ $P_c^*(\frac{3}{2}^{-})$ 
requires $70\,\%$ the attraction of a standard $\Sigma_c \bar{D}^*$ $P_c^*$
to bind (for the $\Sigma_c \Sigma_c \pi$
axial coupling we use $g_3 = -1.38$~\cite{Cheng:2015naa}).
For the heavy baryonium the numbers are $46\,\%$ ($0^{-}$) and 
$53\,\%$ ($1^{-}$) respectively.
The probability of binding is enhanced
but dependent on unknown short-range physics.

If binding happens for distances in which the present picture is valid,
short-range physics will not be necessary.
The radius below which the $P_c^*(\frac{3}{2}^{-})$
binds
is
$0.94\,{\rm fm}$ 
while for the $0^{-}$ / $1^{-}$ baryonia we have
$0.40\,{\rm fm}$ 
/ $0.84\,{\rm fm}$ 
respectively.
For the $1^{+}$ $\Sigma_c \Xi_b'$-$\Lambda_{c1} \Xi_b$ molecule we have
$0.87\,{\rm fm}$ instead.
For $r < 1/2m_{\pi}$ ($\sim 0.7\,{\rm fm}$) two-pion exchange
and hadron finite-size effects dominate,
setting the limits of the OPE description
and providing a criterion for binding.
From this we can be confident about the existence of the $1^{-}$ baryonium and
the $1^{+}$ $\Sigma_c \Xi_b'$-$\Lambda_{c1} \Xi_b$ molecule,
while the $0^{-}$ baryonium is contingent on the unknown short-range physics.
But the more interesting cases are that of the $\Sigma_c D^*$ 
$(\frac{1}{2}^{+})$ /
$\Sigma_c \bar{\Xi}_b'$ 
$(0^{+})$ which bind in p-wave.
Here the vector force effectively induces the existence of a channel
behaving much like an s-wave.
For the $1/2^{+}$ $\Sigma_c D^*$-$\Lambda_c D$ system binding happens
for $0.92\,{\rm fm}$ while for the  $0^+$ baryonium
we have $0.86\,{\rm fm}$, 
radii which point towards the existence of these states.
The bottom-line is that the vector force induces a series of
binding mechanisms which do not require the ratio $m_{\pi} / \mu_{\pi}$
to be particularly large (a factor of $2-3$ is probably enough)
and which in a few cases lead to predictions of new molecules.

Scale invariant hadron molecules are an intriguing theoretical possibility.
They are the two-body realization of a type of universality
that is usually only found in three-body atomic and nuclear systems.
There are clear theoretical requirements for a hadron molecule
to show scale invariance at long distances,
where the most natural mechanism is the exchange of a pion almost
on the mass shell between initial and final two-hadron states
with opposite intrinsic parities.
If we consider heavy hadrons, the candidates include 
$\Lambda_{c1} \bar{D}$-$\Sigma_{c} \bar{D}^*$, i.e.
the molecular interpretation of the recently discovered
$P_c^*$ pentaquark state,
while the most likely scale invariant molecule is probably
the $\Lambda_{c1} \bar{\Xi}_{b}$-$\Sigma_{c} \bar{\Xi}_{b}'$ baryonium.
Discrete scale invariance requires that the couplings have a minimal strength,
a condition that a $\frac{1}{2}^+$ heavy pentaquark and a
$\Lambda_{c1} \bar{\Xi}_{b}$-$\Sigma_{c} \bar{\Xi}_{b}'$
molecule can meet.
The same ideas apply to the $\Lambda_{c1} D$-$\Sigma_{c} D^*$ and
$\Lambda_{c1} {\Xi}_{b}$-$\Sigma_{c} {\Xi}_{b}'$ molecules as
the vector force attraction is independent of
whether we have hadrons or antihadrons.
The appearance of a geometrical spectrum actually requires
the effective mass of the pion to be considerably smaller
than the other hadronic scales in the molecule.
This condition is not likely to be met in nature,
but could very well be realized in the lattice.
Even if there is no geometrical spectrum in these molecules,
the long-range attraction provided by the vector force
plays an important role as a binding mechanism, which
cannot be ignored and in a few cases guarantees binding.
The vector force is indeed a new type of long range dynamics
that has not been previously considered
neither in the $P_c^*$ pentaquark
nor in other hadronic molecules where it is present and can be relevant.
An illustrative example is the enhancement of P-wave
interactions, as happens in the $\frac{1}{2}^+$ $\Sigma_c D^*$
system after we include the $\Lambda_{c1} D$ channel.
In this type of hadronic molecule the role of scale invariance
is analogous to that in the triton, $^4{\rm He}$,
a few halo nuclei and a series of cold atoms systems, to name a few examples.

\section*{Acknowledgments}

This work is partly supported by the National Natural Science Foundation
of China under Grants No. 11375024,  No.11522539 and the Fundamental
Research Funds for the Central Universities.

%

\end{document}